\documentclass[12pt,a4paper,twoside]{article}
\usepackage{epsfig}
\usepackage{graphics}
\newlength{\figwidth}
\newlength{\figheight}
\def\z0{Z}
\def\eg{{e.g.} }
\def\ie{{i.e.} }
\title{
       \vspace{-1.5cm}
       \begin{flushright}
       \begin{tabular}{l}
       {\large CERN-TH/2002-166 }    \\[-3mm]
       {\large IFT - 29/2002}    \\[-3mm]
       {\large hep-ph/0208234}\\
       {\large July 2002}
       \end{tabular}
       \end{flushright}
       \vspace{1.5cm}
      \sc  
The~SM~Higgs-boson~production \\ 
in~\( \gamma \gamma \rightarrow h\rightarrow b\overline{b} \) \\
   at~the~Photon~Collider~at~TESLA}
 \author{Piotr Nie\.zurawski, Aleksander Filip \.Zarnecki \\
 {\small\it Institute of Experimental Physics, Warsaw University, 
    ul. Ho\.za 69, 00-681 Warsaw, Poland} \\[3mm]
 Maria Krawczyk \\
{\small\it Theory Division, CERN, CH-1211 Geneva 23, Switzerland} \\[-2mm]
 {\small and} \\[-2mm]
 {\small\it Institute of Theoretical Physics, Warsaw University, 
        ul. Ho\.za 69, 00-681 Warsaw, Poland} \\[-2mm]
 } 

\date{}

\begin{document} 

\maketitle 

\vfill

\begin{abstract}
 Measuring the \( \Gamma (h\rightarrow \gamma \gamma ){\rm {Br}}
(h\rightarrow b\overline{b}) \) decay
at the photon collider at TESLA is studied for a Standard Model 
Higgs boson of mass \( m_{h}=120 \) GeV. The main background due to the 
process \( \gamma \gamma \rightarrow Q\overline{Q}(g) \), where \( Q=b,\, c \),
is estimated  using the NLO QCD program (G.~Jikia); the results obtained 
are compared with 
the  corresponding LO estimate. Using a realistic luminosity spectrum and 
performing a detector simulation with the SIMDET program, we find that 
the \( \Gamma (h\rightarrow \gamma \gamma ){\rm Br}(h\rightarrow b\overline{b}) \)
decay can be measured with an accuracy better than 2\%
after one year of photon collider running.
% after one year of accelerator running, corresponding
% to the luminosity 
% \( L_{\gamma \gamma }(W_{\gamma \gamma }>80\textrm{ GeV})=
%                               84\textrm{ fb}^{-1} \).}
\end{abstract}

\vfill

\begin{flushleft} 
CERN-TH/2002-166 \\
IFT - 29/2002    \\
hep-ph/0208234 \\
July 2002
\end{flushleft}

% \vfill\eject

\thispagestyle{empty}

\clearpage

\section{Introduction}

A  search of the last missing member of the Standard Model (SM) family,
the Higgs boson, is among the most important tasks for the present and
future colliders. Once the Higgs boson is discovered, it will be crucial
to determine its properties with a high accuracy.  
%in order to confirm its origin. 
A photon collider option of the TESLA
$e^+e^-$ collider \cite{TDR} offers a  unique possibility to produce
the Higgs boson as an \( s \)-channel resonance. The neutral Higgs boson
couples to the photons  through a loop 
with  the massive charged particles. This loop-induced $h\gamma \gamma$ coupling
%As a result the Higgs cross section
is sensitive to contributions of new particles, which appear in various 
extensions of  the SM. 

The SM Higgs boson with a  mass below \( \sim 140\) GeV is expected
to decay predominantly into the \( b\bar{b} \) final state. 
Here we consider the process \( \gamma \gamma \rightarrow h\rightarrow 
b\overline{b} \) for a Higgs-boson mass of \( m_{h}=120 \) GeV
at a photon collider at TESLA.
Both the signal and  background
events are generated according to a  realistic photon--photon luminosity
spectrum \cite{V.Telnov}, parametrized by a CompAZ model \cite{CompAZ}. 
Our  analysis incorporates a  simulation
of the detector response according to the program SIMDET \cite{SIMDET}.

\section{Photon--photon luminosity spectrum}

The Compton backscattering of a laser light off  high-energy
electron beams is considered as a  source of high energy, highly polarized
 photon beams  \cite{Ilya}. 
A simulation of  the  realistic \( \,\, \gamma \gamma  \) luminosity spectra 
for the photon collider at TESLA,
taking into account non-linear corrections and higher order QED processes, 
has become available recently \cite{V.Telnov}.
In this simulation, according to the  current design \cite{TDR}, 
the energy
of the laser photons \( \omega _{L} \) is assumed to be fixed
 for all considered electron-beam energies. 
%
% Present  analysis is based on the CompAZ 
% parametrization \cite{CompAZ} of the spectrum given in \cite{V.Telnov}.

In  the analysis we use the CompAZ parametrization \cite{CompAZ} of 
the spectrum \cite{V.Telnov} to generate  energies of the colliding photons.
We assume that the energy of primary electrons
% equal to \( \sqrt{s_{ee}} \)=210 GeV. This energy 
can be  adjusted in order to  enhance the signal.
Our signal production of a scalar particle corresponds to  the case 
where projection of the total $\gamma \gamma$ 
angular momentum 
on a collision ($z$) axis  \( J_{z}\) is equal to zero.
For  \( \sqrt{s_{ee}} =  2 E_{e} \) = 210 GeV,  we obtain a
 peak of the \( J_{z} \) = 0 component 
of the photon--photon luminosity spectra 
at the invariant mass of the two colliding photons 
\( W_{\gamma \gamma } \) equal to the considered 
 mass of the Higgs boson, \ie \( m_{h} \) = 120 GeV.

\begin{figure}[t]
{\centering \resizebox*{!}{\figheight}%
        {\includegraphics{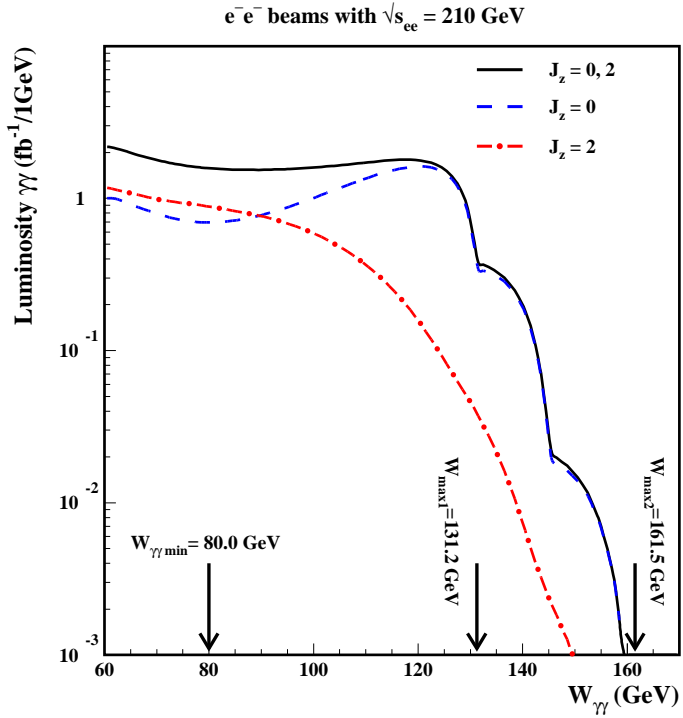}} \par}

\caption{\label{fig:LumiCompAZEee210GeV}Photon--photon luminosity spectra
for \protect\( \sqrt{s_{ee}}=210\protect \) GeV, obtained with CompAZ
parametrization of  Telnov's simulation, as a function of the
invariant mass of two colliding photons $ W_{\gamma \gamma }$.
The contributions of states with the total $\gamma \gamma$ angular momentum 
projected  on a collision ($z$) axis, $J_{z}$=0 and $J_{z}=\pm 2$ 
(denoted as \( J_{z}=2 \)),
are shown separately.
}
\end{figure}

The luminosity spectra are shown in Fig. \ref{fig:LumiCompAZEee210GeV}.
The lowest invariant mass of the two colliding photons
used  in our generation of events
is equal to \( W_{\gamma \gamma \min } = 80 \) GeV. 
For the assumed \( \sqrt{s_{ee}} \) value, 
the maximum invariant mass for the colliding photons, each  produced in a single 
Compton scattering, is equal to \( W_{\max 1} \) = 131.2 GeV. 
However, there is also
a small contribution from the  events, which correspond to the
interaction of an initial electron with  two laser photons 
(higher order effect). This  gives a higher  maximal invariant mass of 
the produced energetic photon beams, namely 
\( W_{\max 2} \) = 161.5 GeV.
The results presented in this paper were obtained 
for an integrated luminosity
of the primary $e^- e^-$ beams equal to \( L_{ee}^{geom} =502 \; \rm fb^{-1} \), 
expected for one year of the photon collider running \cite{V.Telnov}.
The resulting $\gamma \gamma$ luminosity is then expected to be: 
\( L_{\gamma \gamma } = 409 \;\rm fb^{-1} \),
or  84 \( \rm fb^{-1} \)  for $W_{\gamma \gamma }>80$ GeV .

\begin{figure}[htb]
% \vspace{-2cm}
{\centering \resizebox*{!}{\figheight}%
         {\includegraphics{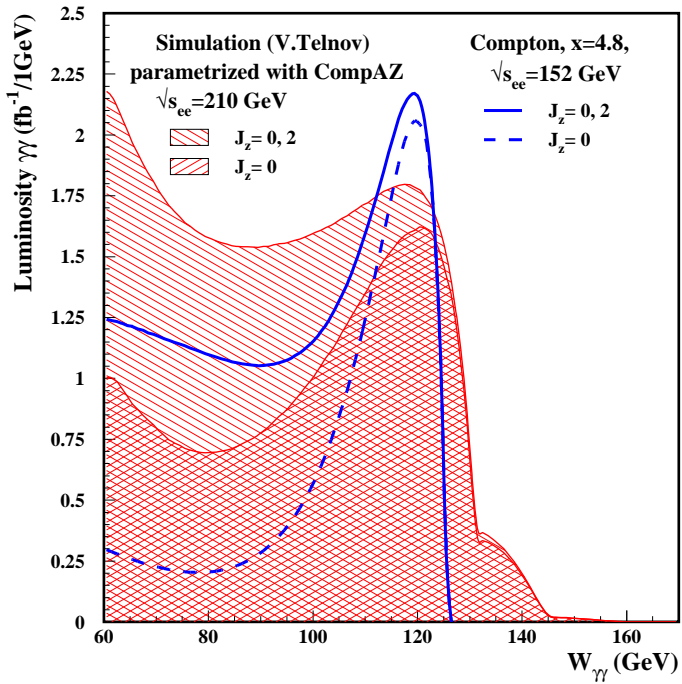}} \par}

\caption{\label{fig:LumiComparison}Photon--photon luminosity spectra used
in the analysis of the SM Higgs-boson production with mass $ m_{h}=120$ GeV,
as a function of the invariant mass of two colliding 
photons $ W_{\gamma \gamma }$.
The spectrum used here, 
as obtained from CompAZ parametrization 
based on Telnov simulation
% (center-of-mass energy of electron beams $ \sqrt{s_{ee}}$=210 GeV), 
(hatched areas), is compared with a spectrum derived from the lowest order QED predictions 
for the Compton scattering, used in the earlier analysis (lines). 
% ($\sqrt{s_{ee}}$=152 GeV).
%
The total luminosity distribution ($J_{z}=0, \pm 2$) 
and the $J_{z}=0$ contribution are shown, separately.
}
\end{figure}

In the earlier analyses, for instance in \cite{JikiaAndSoldner},
the spectra that were used had been derived from the 
lowest order QED calculation for the
Compton scattering,  with a fixed 
parameter \( x=\frac{4E_{e}\omega _{L}}{m_{e}^{2}} \)
equal to 4.8. 
%independently of a energy of the electron-beam, \( E_{e} \). 
%
The realistic spectrum \cite{V.Telnov}, parametrized by the CompAZ model, 
differs significantly from the spectrum of the high-energy photons 
used in \cite{JikiaAndSoldner}, which  is shown 
in~Fig. \ref{fig:LumiComparison}.
%
%Significant 
%differences between the realistic spectrum  % with varying $x$,
%and the lowest order Compton formula
%used in the earlier analysis can be seen. 
%
%Contributions of
The comparison is made for two 
combinations of the  helicities of two colliding photons, 
\( (\pm ,\pm ) \) and \( (\pm ,\mp ) \),  
with the total angular momentum projected on a collision ($z$) axis 
 equal to 0 and $\pm 2$, respectively.

\section{Details of a simulation and the first results }

We calculate the total width and  branching ratios of the SM Higgs
boson, using the program HDECAY \cite{HDECAY}, where
higher order QCD corrections are included. 
A generation of events was done with
the PYTHIA 6.205 program \cite{PYTHIA}, with the parameters for a Higgs boson
as in the  HDECAY. 
A parton shower algorithm, implemented in PYTHIA,
was used to generate the final-state particles. 

The  background events due to processes 
\( \gamma \gamma \rightarrow b\bar{b}(g),\, \, c\bar{c}(g) \)
were  generated using the program written by G.~Jikia \cite{JikiaAndSoldner},
where a complete  NLO QCD  calculation for the production of  massive
quarks is performed within the massive-quark scheme. 
The program includes exact one-loop QCD corrections to the lowest order
(LO) process
\( \gamma \gamma \rightarrow b\bar{b},\, \, c\bar{c} \)
\cite{JikiaAndTkabladze}, and in addition 
the non-Sudakov form factor in the double-logarithmic
approximation, calculated up to four loops \cite{MellesStirlingKhoze}.

For a comparison we generate also the  LO background events,
using the QED Born cross section for the processes
 \( \gamma \gamma \rightarrow b\bar{b} \) and \( \gamma \gamma \rightarrow c\bar{c} \), 
 including in addition a parton shower, as  implemented in  PYTHIA.\footnote{%
For  consistency with  Jikia's program, we use a fixed 
electromagnetic coupling constant equal to 
\( \alpha _{em} \approx 1/137 \). 
}

The fragmentation into hadrons was performed using the PYTHIA program. A fast
simulation  for a TESLA detector, the 
 program  SIMDET version 3.01 \cite{SIMDET},
was used  to model a detector performance. The jets were reconstructed
using  the Durham algorithm, with \( y_{cut} = 0.02 \); the distance measure
was defined as 
\( y_{ij}=2\min (E^{2}_{i},E^{2}_{j})(1-\cos \theta _{ij})/E^{2}_{vis} \),
where $E_{vis}$ is defined as the total energy measured in the detector.

The double $b$-tag was required to select the signal
\( h\rightarrow b\bar{b} \) events. Since no suitable flavour-tagging 
package exists
%simulation 
%is not implemented 
for the SIMDET 3.01 program\footnote{%
The $b$-tagging code adapted for the new version of SIMDET 4.01 \cite{SIMDET401} 
should become available  soon \cite{Btagging}.
%The implementation of $b$-tagging for new version of SIMDET 4.01 \cite{SIMDET401} 
%should become publicly available  soon \cite{Btagging}.
}%
, we assume, following  the approach used in \cite{JikiaAndSoldner},
 a fixed efficiency for the $b\bar{b}$-tagging, equal to
  \( \varepsilon _{bb}=70\% \), and a fixed 
 probability  for a mistagging of the 
\( c\bar{c} \) events, a main background to the \( b\bar{b} \) events, 
equal to
\( \varepsilon _{cc}=3.5\% \). 

\begin{figure}[htb]
{\centering \resizebox*{!}{\figheight}%
            {\includegraphics{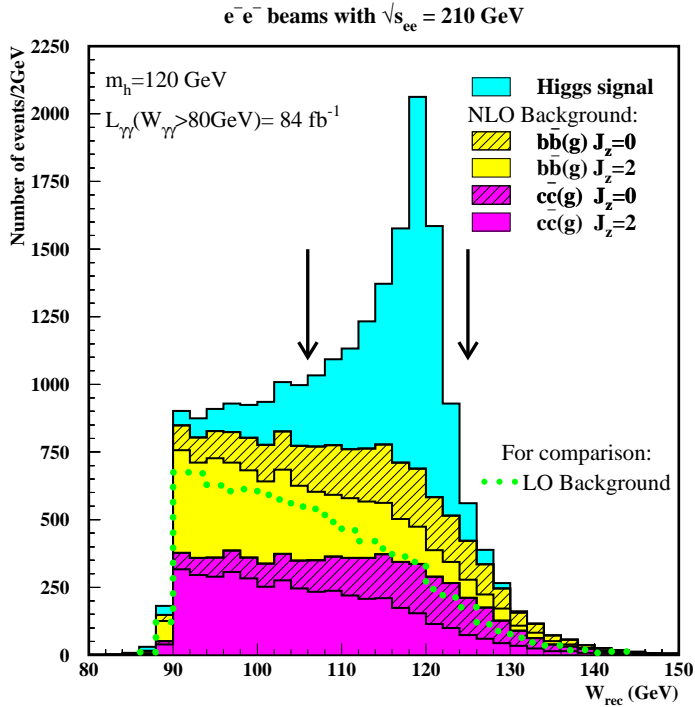}} \par}

\caption{\label{fig:ResultWithNLOBackgd}
Reconstructed invariant mass \protect\( W_{rec}\protect \)
distributions for the selected $b\bar{b}$ events.
Contributions of the signal, due to the Higgs boson with a mass 
$m_h = 120$ GeV, and of the heavy-quark 
 background, calculated in the NLO QCD, are indicated. For
 comparison, the LO background estimate  is also plotted (dots). 
Arrows indicate the mass window optimized for the measurement of the 
\( \Gamma (h\rightarrow \gamma \gamma ){\rm Br}(h\rightarrow b\overline{b}) \).
}
\end{figure}

The following cuts were used  to 
%suppress a background:
select properly  reconstructed $b \bar{b}$ events:
\begin{enumerate}
\item a total visible energy \( E_{vis} \) greater than 90 GeV,

\item since the Higgs boson is expected to be produced
  almost at rest, the ratio of the total longitudinal momentum of all 
 observed particles to the total visible energy is taken to be 
\( |P_{z}|/E_{vis}<0.1 \),

\item a number of jets \( N_{jets}=2,\, 3 \), so that 
   events with one additional jet due to hard-gluon emission are accepted,

\item for each jet, $i=1, ..., N_{jets}$, we require 
   \( |\cos \theta _{i}|<0.75 \),  where
\( \cos \theta _{i}=p_{z\, i}/|\overrightarrow{p_{i}}| \).

\end{enumerate}
Using the above cuts, we obtain the
 distributions of the reconstructed \( \gamma \gamma  \) invariant
mass \( W_{rec} \) for a signal and for a background,  shown in 
Fig. \ref{fig:ResultWithNLOBackgd}. The  $J_z=0$ and  $J_z= \pm 2$
contributions from the  NLO background,
with $bb(g)$ and with $cc(g)$ final states,  
are shown separately. For a comparison, the estimated LO background 
is presented as well (dotted line). 
Note that the NLO background contribution
is  approximately two times larger than the LO one.
This is mainly due to the $J_z=0$ component, 
which is strongly suppressed in the LO case; however in the case of NLO it gives
a large contribution,
especially pronounced in the high-$W_{\gamma \gamma }$ part of the signal peak.
A more detailed comparison of the LO and NLO background estimations
 is presented in Appendix. 
Other background contributions, from the resolved photon(s)
 interactions and the overlaying events, were found to be negligible \cite{Overlay}.

%On the parton level we studied also other background sources: resolved photon(s)
% interactions and overlaying events \cite{Overlay}. Both contributions are found to be negligible.

\section{Final results}

Assuming that the signal for  Higgs-boson production  
will be extracted
by counting the \( b\bar{b} \) events in the mass window around the peak, 
% with a window width corresponding to a Higgs-boson mass resolution,  
and subtracting the background events expected in this window,
we can calculate 
the expected relative statistical error for  the partial width multiplied
by the branching ratio,
\( \Gamma (h\rightarrow \gamma \gamma ){\rm Br}(h\rightarrow b\overline{b}) \),
in the following way:
\[
\frac{\Delta \left[ 
\Gamma (h\rightarrow \gamma \gamma ){\rm Br}(h\rightarrow b\overline{b})\right] }
{\left[ \Gamma (h\rightarrow \gamma \gamma ){\rm Br}(h\rightarrow b\overline{b})
 \right] }=\frac{\sqrt{N_{obs}}}{N_{obs}-N_{bkgd}}\; .\]
The accuracy expected for the considered quantity
\( \Gamma (h\rightarrow \gamma \gamma ){\rm Br}(h\rightarrow b\overline{b}) \),
if estimated from the reconstructed invariant-mass distribution
obtained for the  Higgs-boson mass of 120 GeV in the selected
mass region, between 106 and 126 GeV 
(see  Fig. \ref{fig:ResultWithNLOBackgd}),
is equal to 1.9\%. It is   in agreement with 
the result of a previous analysis \cite{JikiaAndSoldner}.

\begin{figure}[t]
{\centering \resizebox*{!}{1.2\figheight}%
           {\includegraphics{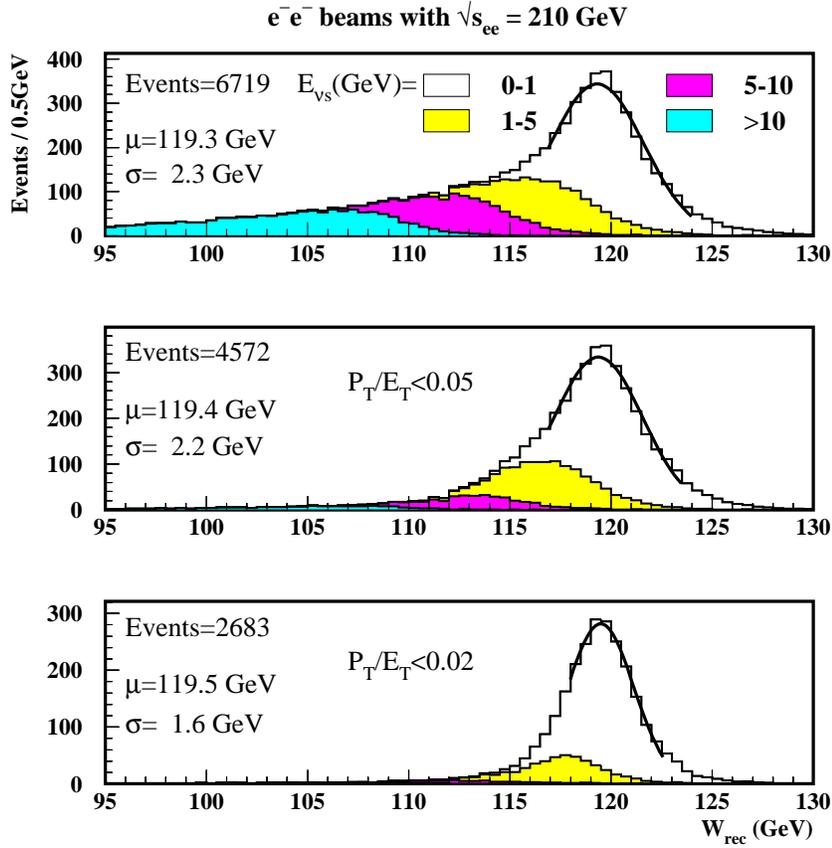}} \par}

\caption{\label{fig:PTNeutrinos}
Reconstructed invariant mass $ W_{rec}$ distributions 
for $\gamma \gamma \rightarrow h \rightarrow b \bar{b}$ events,
for various $P_{T}/E_{T}$ cuts.
Contributions of events with a different total energy 
of neutrinos in the event $ E_{\nu s}$
are indicated by different colours. 
The parameters $ \mu $ and $\sigma$ are obtained from the Gaussian fit 
in the region $ (\mu -\sigma ,\, \, \mu +2\sigma )$.
}
\end{figure}

A long tail in the reconstructed mass \( W_{rec} \) distribution obtained   
for the \( h\rightarrow b\bar{b} \) events seen in 
Fig. \ref{fig:ResultWithNLOBackgd} is due to the 
escaping neutrinos, which  mainly originate in the semileptonic decays of 
the $D$-  and $B$-mesons.
This tail  can be effectively suppressed
by applying an additional cut   on \( P_{T}/E_{T} \), where
\( P_{T} \) and \( E_{T} \) are the absolute values of the total transverse
momentum of an event $\vec{P}_{T}$ and the total transverse energy, 
respectively.\footnote{$\vec{P}_{T}$ ($E_{T}$) is calculated as
a vector (scalar) sum of the transverse momenta 
$\vec{p}_T^{\;i} = (p^i_x, p^i_y, 0)$
(the transverse energies $E_T^i = E^i \sin \theta_i $)
over all particles that belong to  an event.} The cut relies on demanding the \( P_{T}/E_{T}\) 
to be small.
The \( W_{rec} \) distributions for the 
 \( h\rightarrow b\bar{b} \) events obtained by applying  various 
 \( P_{T}/E_{T} \) cuts are shown in
Fig. \ref{fig:PTNeutrinos}. In this figure we use different colours to denote
the different  total energies of neutrinos in the event, \( E_{\nu s} \). 
The effects due to the detector resolution  
influence a shape of the distribution for \( W_{rec} > m_{h} \),
whereas for a lower \( W_{rec} \) one   observes
a significant effect  of the escaping neutrinos
in the distribution. 
By applying a realistic \( P_{T}/E_{T} \, \) cut, \eg \( P_{T}/E_{T} < 0.04 \), 
we can obtain a mass resolution, derived from the 
Gaussian fit in the region from \( \mu -\sigma  \)
to \( \mu +2\sigma  \),  better than 2 GeV.

\begin{figure}[htb]
{\centering \resizebox*{!}{\figheight}%
             {\includegraphics{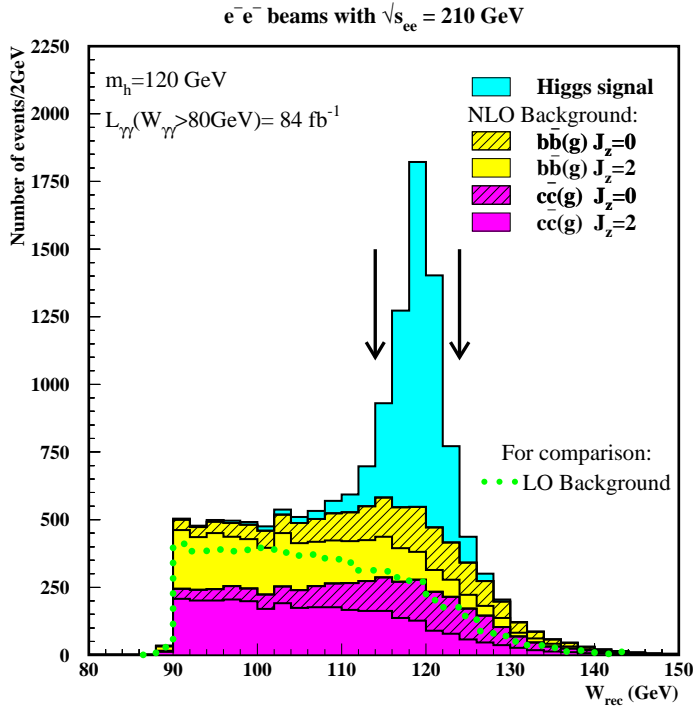}} \par}

\caption{\label{fig:WrecWithPToverETlt0.04}
As in Fig. \ref{fig:ResultWithNLOBackgd},
 for the reconstructed invariant mass \protect\( W_{rec}\protect \)
distributions for the selected $b\bar{b}$ events, obtained by applying an 
additional cut on the ratio of the total transverse momentum and
total transverse energy, $P_{T}/E_{T} < 0.04$.
}
\end{figure}

Shown in Fig.~\ref{fig:WrecWithPToverETlt0.04} is
the \( W_{rec} \) distribution obtained by
applying the cut  \( P_{T}/E_{T}<0.04 \).
The relative accuracy expected
for the \( \Gamma (h\rightarrow \gamma \gamma ){\rm Br}(h\rightarrow b\overline{b}) \)
measurement,
calculated in the \( W_{rec} \) mass range between 114 and 124 GeV
(as indicated by arrows in the figure), is equal to 2.2\%. 
We conclude that  the  \( P_{T}/E_{T} \) cut improves 
a mass resolution, but  worsens  the statistical significance of the measurement.

\begin{figure}[htb]
% \vspace{-1cm}
{\centering \resizebox*{!}{\figheight}%
               {\includegraphics{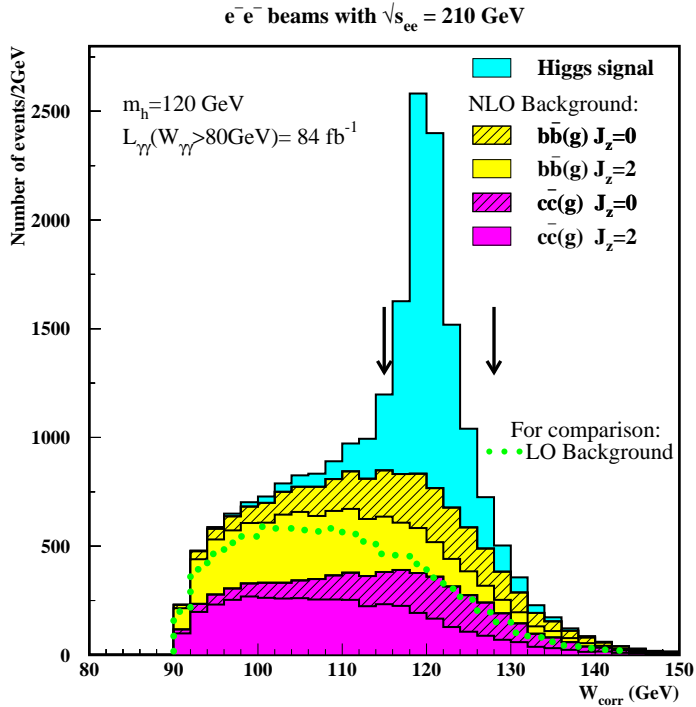}} \par}

\caption{\label{fig:Wcorr}
As in Fig. \ref{fig:ResultWithNLOBackgd}, for the corrected invariant mass \protect\( W_{corr}\protect \)
distributions.
}
\end{figure}

We have found a method,  which allows an increase
of a signal-to-background ratio without reducing the event statistics. 
We  assume that the measured missing transverse
momentum is due to a single neutrino emitted perpendicularly to the
beam line.\footnote{%
Due to a large spread of the photon beam energy, no constraints can
be imposed on the longitudinal momentum.}
Then, we introduce  the corrected, reconstructed invariant mass as: 
\begin{equation}
\label{eq:Wcorr2}
W_{corr}\equiv \sqrt{W^{2}_{rec}+2P_{T}(E_{vis}+P_{T})},
\end{equation}
The distributions of the \( W_{corr} \), obtained for the  signal and
background events,  are shown in Fig.~\ref{fig:Wcorr}. 
The most precise measurement of the Higgs-boson production cross section
is obtained using the  mass window \( W_{corr} \)
between 115 and 128 GeV, as indicated by arrows.
In the selected \( W_{corr} \) region one expects, after one year of
photon collider running at nominal luminosity,
about 5900 reconstructed signal
events and 4600 background events  (\ie \( S/B \approx 1.3\)).
This corresponds to the expected relative 
statistical precision of  the measurement:
%
% Varmin=115. Varmax=128. Precision with LO= 1.5447 % with NLO=1.73505 %
%
\[
\frac{\Delta \left[ 
\Gamma (h\rightarrow \gamma \gamma ){\rm Br}(h\rightarrow b\overline{b})\right] }%
{\left[ \Gamma (h\rightarrow \gamma \gamma ){\rm Br}(h\rightarrow b\overline{b})
         \right] }=1.7\% \; . \]
%
% This is the final result of our analysis.

%\pagebreak

\section{Conclusions}

Our analysis shows that, for the SM Higgs boson with a mass around 120 GeV,
the two-photon width for the \( b\bar{b} \) final state can be measured
in the photon collider at TESLA with a precision better than 2\%.
If the reconstructed invariant mass of the event is corrected
for the energy of escaping neutrinos from $D$- and $B$-meson decays,
we achieve a precision of the measurement of the 
\( \Gamma (h\rightarrow \gamma \gamma ){\rm Br}(h\rightarrow 
b\overline{b}) \) equal to  1.7\%. 
The obtained accuracy
is in an agreement with the result of a previous analysis, based on the
idealistic  Compton spectrum \cite{JikiaAndSoldner}. 
Note, however, that the  
 realistic photon--photon luminosity spectrum  that we use
is more challenging.
% for a precision measurement of the Higgs boson width 
%\( \Gamma (h\rightarrow \gamma \gamma ) \).
%

The measurement discussed in this paper can be used to derive the partial width
\( \Gamma (h\rightarrow \gamma \gamma ) \), taking 
\( {\rm Br}(h\rightarrow b\overline{b}) \) value from precise 
measurement at the $e^+ e^-$ Linear Collider. 
With 1.7\% accuracy on 
\( \Gamma (h\rightarrow \gamma \gamma ){\rm Br}(h\rightarrow b\overline{b}) \),
obtained in this analysis, assuming \( {\rm Br}(h\rightarrow b\overline{b}) \) 
will be measured to 1.5\% \cite{Brient}, Higgs-boson partial width 
\( \Gamma (h\rightarrow \gamma \gamma ) \) can be extracted with accuracy of 2.3\%.
Using in addition the result from the $e^+ e^-$ Linear Collider for 
\( {\rm Br}(h\rightarrow \gamma \gamma) \) 
%to 10\% 
\cite{Boos}, one can also extract 
\( \Gamma_{\rm tot} \) with precision of 10\%.

% From the measurement of \( \Gamma (h\rightarrow \gamma \gamma ){\rm Br}(h\rightarrow 
% b\overline{b}) \)  to  1.7\% one can derive the parameter 
% \( \Gamma (h\rightarrow \gamma \gamma ) \) to 2.9\% (2.3\%), 
% by using the information on the expected accuracy 2.4\% (1.5\%) for 
% \( {\rm Br}(h\rightarrow b\overline{b}) \) at the $e^+ e^-$ Linear Collider \cite{Brient}.
% Using in addition the result from the $e^+ e^-$ Linear Collider for 
% \( {\rm Br}(h\rightarrow \gamma \gamma) \) to 10\% \cite{Boos}, one can get the basic 
% quantity \( \Gamma_{\rm tot} \) with precision of 10\%.

%We plan to extend the presented 
The SM Higgs-boson production
$\gamma \gamma \rightarrow h \rightarrow b\bar{b}$ can be considered
%in a~forthcoming paper to 
for masses up to about 160~GeV \cite{JikiaAndSoldner}.  For  higher
masses of the SM Higgs boson one should consider other decay channels,
see \eg \cite{wwzz}.

\subsection*{Acknowledgements}

We would like to thank S.~S\"{o}ldner-Rembold for valuable discussions
and for giving us an access to the program written by G.~Jikia, which
generates the NLO background. Also fruitful discussions with I.~Ginzburg
and V.~Telnov are acknowledged. The work was partially sponsored
by the BMBF--KBN collaboration program TESLA. M.K.~acknowledges partial
support by the Polish Committee for Scientific Research, Grants 2 P03B 05119 (2002), 
5 P03B 12120 (2002), and by the European Community's
Human Potential Programme under contract HPRN-CT-2000-00149 Physics
at Colliders.

\section*{Appendix: Comparison of the LO \\ \hspace*{3.3cm} and NLO background estimates }

\begin{figure}[hb]
{\centering \resizebox*{!}{1.0\figheight}%
         {\includegraphics{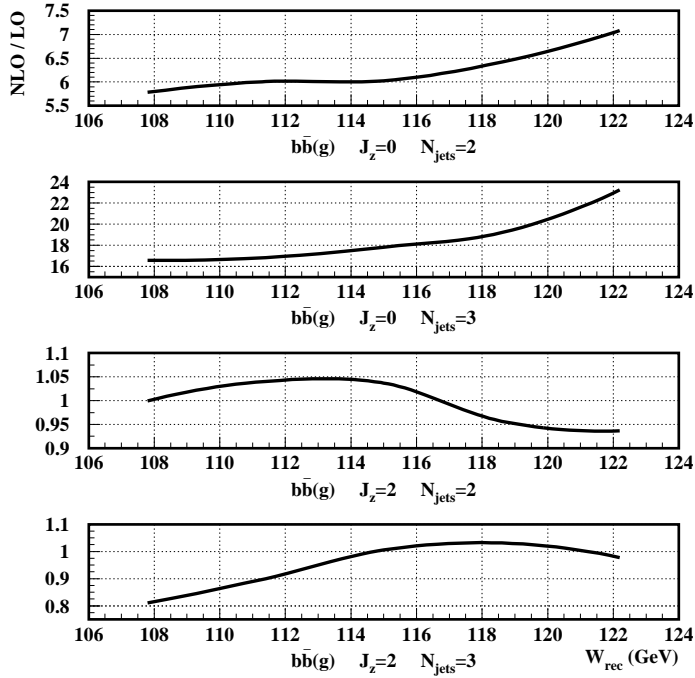}} \par}

\caption{\label{fig:BackgroundNLOoverLOratio}Ratio of the NLO to  LO results
for the reconstructed invariant mass \protect\( W_{rec}\protect \)
distribution for the background process  \protect\( \gamma \gamma \rightarrow b\bar{b}(g)\protect \).
}
\end{figure}

As it is well known, 
see e.g.~\cite{JikiaAndTkabladze, MellesStirlingKhoze},
the NLO corrections for the background process   
\( \gamma \gamma \rightarrow b\bar{b} \) are large. Whereas the LO
background  is strongly suppressed for  a  \( J_{z}=0 \)
contribution, this suppression is removed for the  higher order process
with an additional  gluon in the final state.
As an example, we show in Fig. \ref{fig:BackgroundNLOoverLOratio}
 the  ratios of the corresponding NLO and  LO results for the $W_{rec}$
distribution,  for the 
process \( \gamma \gamma \rightarrow b\bar{b}(g) \), for different
values of \( N_{jets} \) and \( J_{z} \). In each case  one observes
large 
differences between the NLO and LO results, both in the shape and in the
normalization. As the precise determination of the background shape
is crucial for a reliable  estimation of the Higgs-boson  width, we conclude
that a rescaling of the LO estimates cannot be recommended as a
substitute of the full NLO background analysis.

\end{document}